\documentclass[12pt]{article}
\usepackage{graphicx,subcaption}

\newcommand{\NOVA}{NO$\nu$A} 

\newcommand\pubnumber{NuPhys2017-van Eijk}
\newcommand\pubdate{\today}

\def\wipac{WIPAC, UW Madison\\
222 West Washington Ave, 53703, Madison, WI, USA}

\def\Title#1{\begin{center} {\Large #1 } \end{center}}
\def\Author#1{\begin{center}{ \sc #1} \end{center}}
\def\Address#1{\begin{center}{ \it #1} \end{center}}

\newcommand\pubblock{\rightline{\begin{tabular}{l} \pubnumber\\
         \pubdate  \end{tabular}}}
\newenvironment{Abstract}{\begin{quotation}  }{\end{quotation}}
\newenvironment{Presented}{\begin{quotation} \begin{center} 
             PRESENTED AT\end{center}\bigskip 
      \begin{center}\begin{large}}{\end{large}\end{center} \end{quotation}}

\def\beq{\begin{equation}}
\def\eeq#1{\label{#1}\end{equation}}
\def\eeqn{\end{equation}}

\def\beqa{\begin{eqnarray}}
\def\eeqa#1{\label{#1}\end{eqnarray}}
\def\eeqan{\end{eqnarray}}

\let\bar=\overbar

\def\Dslash{\not{\hbox{\kern-4pt $D$}}}
\def\dslash{\not{\hbox{\kern-2pt $\del$}}}

\def\msb{{\bar{\ssstyle M \kern -1pt S}}}

\begin{document}
\begin{titlepage}
\pubblock

\vfill
\Title{Electronics and DAQ for the CHIPS Experiment}
\vfill
\Author{Daan van Eijk}
\Address{\wipac}
\vfill
\begin{Abstract}
CHIPS (CHerenkov detectors In mine PitS ) is a novel neutrino detector concept, aimed at building megaton water-Cherenkov neutrino detectors in a flexible and cheap way, while yielding science results comparable and contributing to conventional long-baseline neutrino experiments. In the summer of 2018, a 5 kiloton proof-of-principle detector will be installed in a disused water-filled mine pit located in the NuMI neutrino beamline path in Minnesota, USA. The submerged cylindrical detector volume is 25 meters in diameter and 10 meter tall and is surrounded by light-tight liners. All inside walls are covered with PMT holding structures. CHIPS will use thousands of 3-inch PMTs to detect neutrinos interacting in the high-purity water in the detector volume. 

The focus of the presentation at the NuPhys2017 conference was on DAQ and electronics for the CHIPS experiment.
\end{Abstract}
\vfill
\begin{Presented}
NuPhys2017, Prospects in Neutrino Physics\\
Barbican Centre, London, UK,  December 20--22, 2017
\end{Presented}
\vfill
\end{titlepage}
\def\thefootnote{\fnsymbol{footnote}}
\setcounter{footnote}{0}

\section{Introduction}\label{sec:introduction}

CHerenkov detectors In mine PitS (CHIPS) is a modular water-Cherenkov detector concept. The goal of CHIPS is to prove that it's possible to study neutrino oscillation physics using a water-Cherenkov detector at a fraction of the costs of present neutrino detectors. Current cost estimates for CHIPS are \$200k/kiloton detector (including location and infrastructure), to be compared with \$2-10M/kt for conventional water-Cherenkov detectors or \$10-20M/kt liquid Ar detectors. 

This presentation will focus on DAQ and electronics for the CHIPS experiment.

\section{The CHIPS Detector}\label{sec:detector}
\begin{figure}[htb]
\centering
\includegraphics[height=3in]{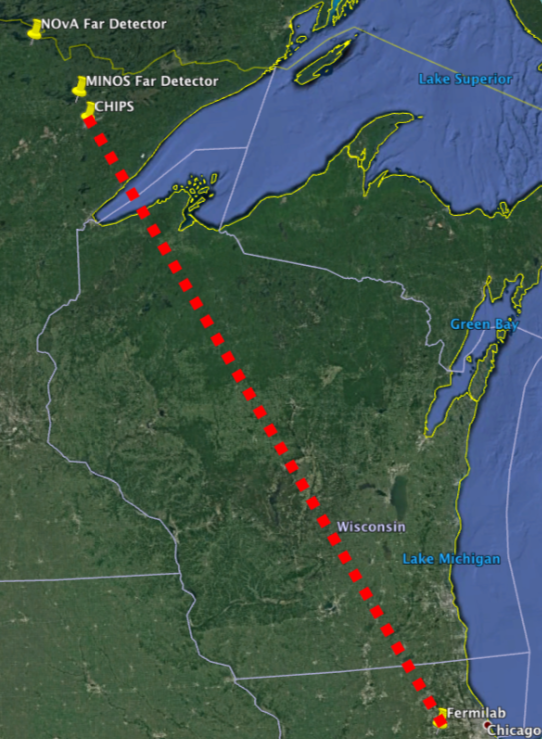}
\caption{Geographic location of the CHIPS proof-of-concept in the NuMI neutrino beam coming from Fermilab. Both the MINOS and \NOVA far detectors are shown as well.}\label{fig:location}
\end{figure}

A 5 kiloton proof-of-principle CHIPS detector is scheduled to be deployed in the summer of 2018 in a flooded mine pit of 60 meters depth in Minnesota (USA), 7 mrad off-axis from the NuMI neutrino beam coming from Fermilab, see Fig.\,\ref{fig:location}. 

\begin{figure}[htb]
    \centering
    \begin{subfigure}[b]{0.49\textwidth}
        \includegraphics[width=\textwidth]{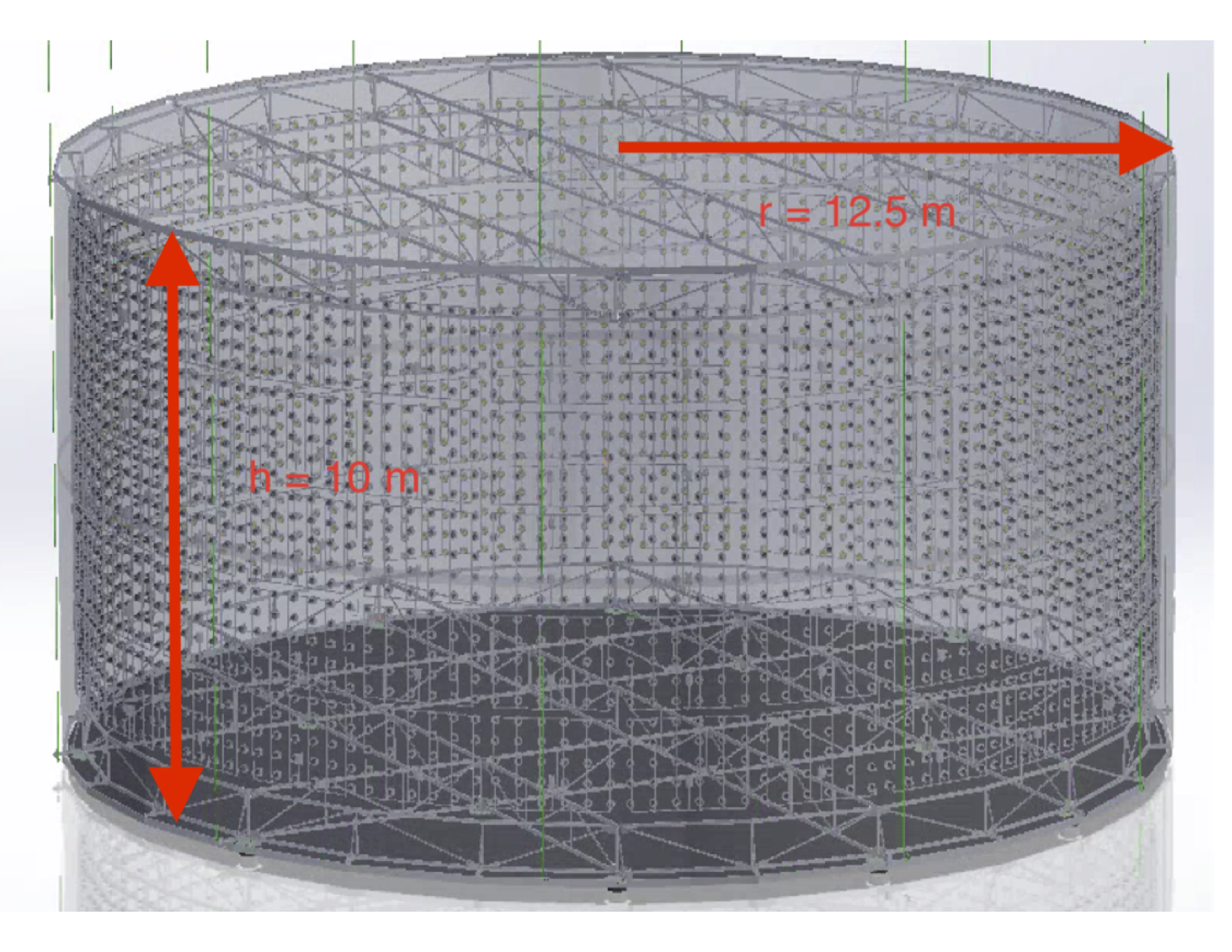}
    \end{subfigure}
    \hfill
    \begin{subfigure}[b]{0.49\textwidth}
        \includegraphics[width=\textwidth]{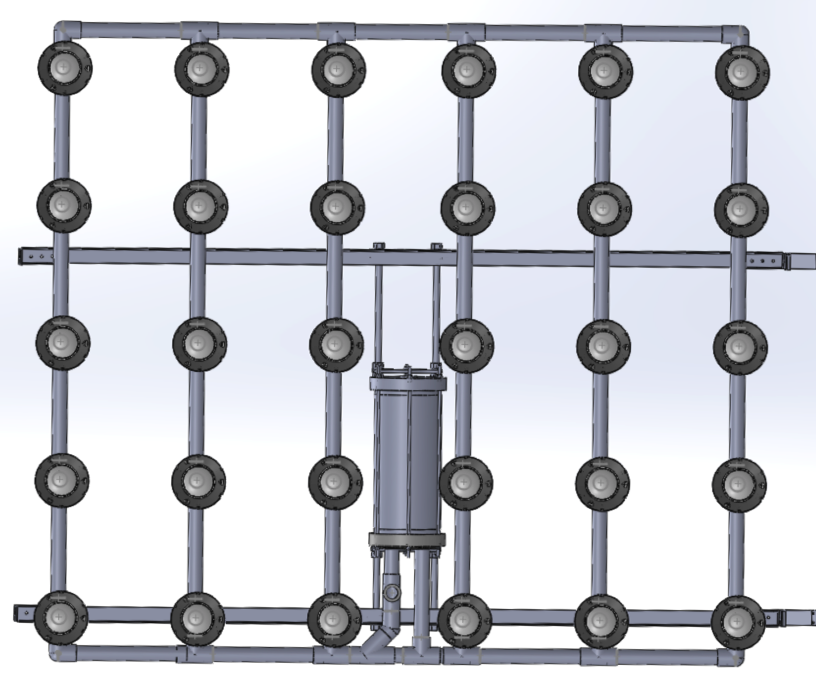}
    \end{subfigure}
    \caption{Schematic picture of the cylindrical proof-of-concept CHIPS detector (left). All inside walls (top, bottom and sides) are covered with detector planes (right) that serve as support structures for the inward-looking 3-inch PMTs.}\label{fig:detector}
\end{figure}

The detector is a cylindrical volume measuring 25 meters in diameter and 10 meters in height. The cylinder outline is formed by a light-tight liner, separating the lake water from filter-circulated water on the inside of the detector. Neutrino interactions in the water produce Cherenkov light, which is detected by so-called detector planes that serve as support structures for the 3-inch photomultiplier tubes (PMTs) that fully cover the inside of the cylinder walls. A schematic picture of the detector is show in Fig.\,\ref{fig:detector}. 

The CHIPS detector will consist of high-density and low-density PMT planes. Both the end caps and the cylinder wall in the forward region (with respect to the neutrino beam) will consist of planes with high PMT coverage of 6\%, while the backward region planes have a 4\% PMT coverage. 

\section{Physics Goals}\label{sec:physics}
\begin{figure}[htb]
\centering
\includegraphics[height=2in]{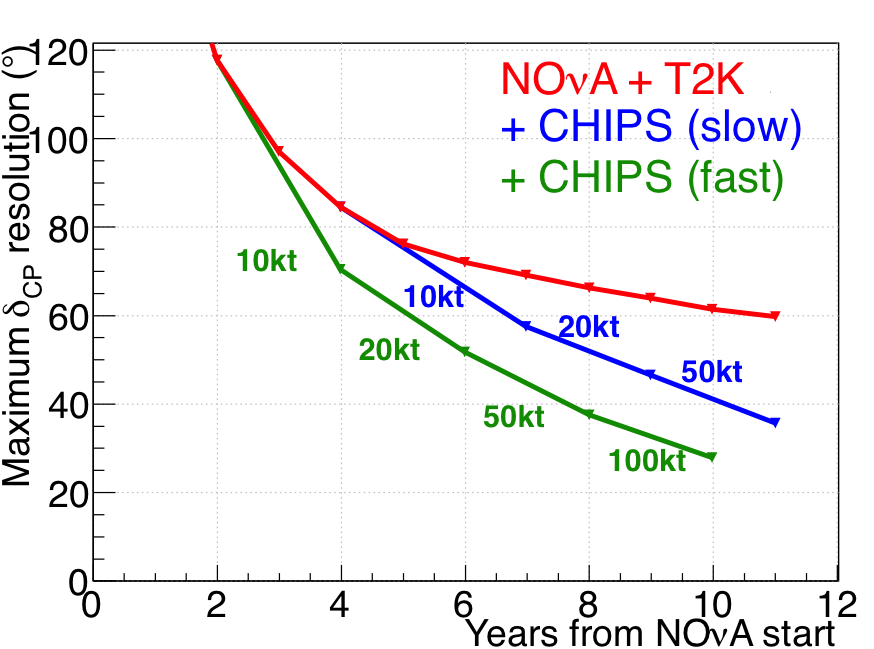}
\caption{Contribution of CHIPS to the measurement of $\delta_{\mathrm{CP}}$ by \NOVA and T2K. Both a slow and a fast development of CHIPS detectors of various detector masses are shown.}\label{fig:deltacp}
\end{figure}

Despite the fact that the total size of any neutrino detector is key, even a moderately sized CHIPS detector of 10-20 kt can significantly contribute to the determination of $\delta_{\mathrm{CP}}$ as performed by \NOVA, see Figure\,\ref{fig:deltacp}. In addition, CHIPS can contribute to precision measurements of $\theta_{13}$ and $\theta_{23}$. 

\section{Electronics and DAQ}\label{sec:electronics}
CHIPS will consist of 6400 PMTs in total. 5500 of these PMTs are Hamamatsu R12199-02, the same type as used in KM3NeT \cite{km3net}. Time-over-threshold (ToT) signals of 30 PMTs that comprise one detector plane are computed on a Central Logic Board (CLB) in a water-tight electronics box mounted on each plane.

\begin{figure}[htb]
\centering
\includegraphics[width=.9\textwidth]{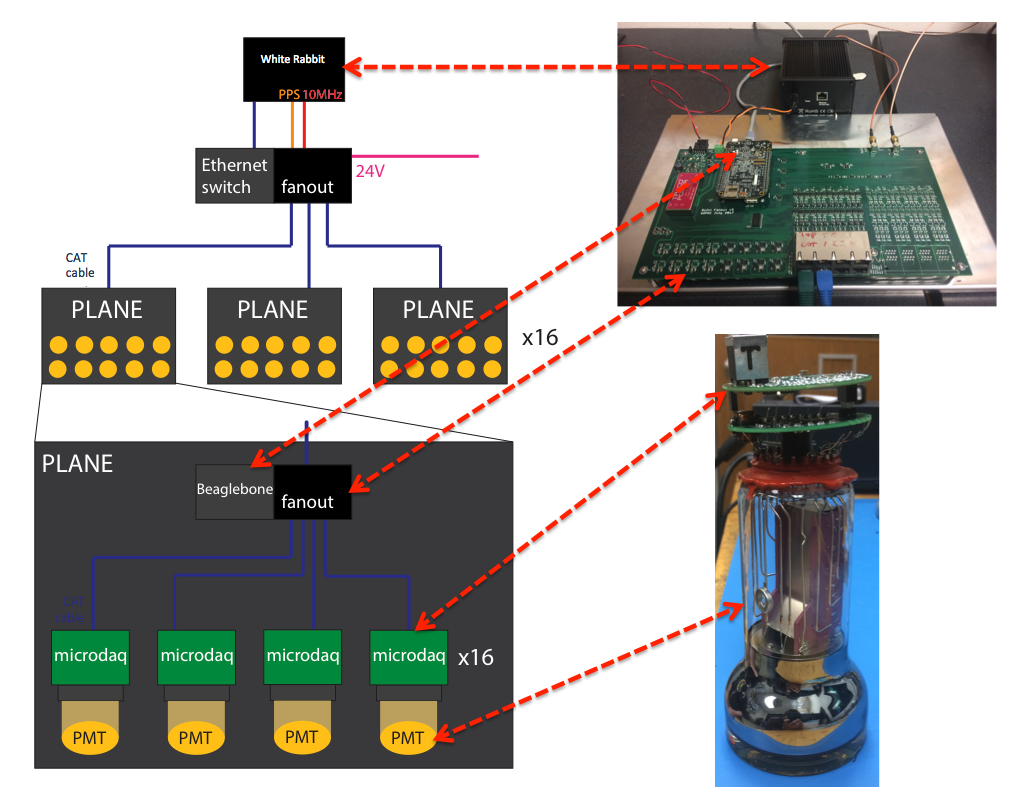}
\caption{MicroDAQ readout scheme for the NEMO3 PMT detector planes (left) and pictures of the various electronics boards (right). From top to bottom: the White Rabbit system, the 16-channel fanout board with the Beaglebone board mounted on it and finally the PMT stack, consisting of the PMT itself, the HV base electronics board and the microDAQ electronics board.}\label{fig:readoutelectronics}
\end{figure}

An additional set of 900 PMTs is refurbished from the NEMO3 experiment \cite{nemo}. For this subset of PMTs an even cheaper electronics readout scheme is being developed. In order to record ToT signals, these PMTs have a microprocessor board called microDAQ connected to them having the same circular form factor as the PMT Cockcroft-Walton base. By using a series of time-delay buffers on the microDAQ board, the full signal shape can also be reconstructed. ToT data from 16 PMTs in one plane are sent to a BeagleBone board \cite{beagleboard} for event building. The BeagleBone communicates with higher level fanout boards through standard ethernet CAT cables. This readout scheme and its various electronics boards are shown in Fig.\,\ref{fig:readoutelectronics}. 

Both the KM3NeT-style readout scheme and the microDAQ readout scheme uses White Rabbit technology \cite{whiterabbit} to provide a 10 MHz clock reference and absolute timing through a so-called pulse-per-second (PPS) signal. Readout of the entire detector is only performed during the 10 $\mu$s spills from the NuMI beam to minimise background events from cosmics. Finally, all data is sent to shore through optical fibers using course wavelength division multiplexing.



\end{document}